\documentclass[a4paper]{article}

\usepackage{INTERSPEECH2021}

\usepackage{times}
\usepackage{latexsym}
\usepackage{multirow}

\usepackage[utf8]{inputenc}
\usepackage{soulutf8}
\usepackage{todonotes}
\usepackage{placeins}
\usepackage{tikz}
\usepackage{svg}
\usepackage{hyperref}

\newcommand\citep{\cite}

\usepackage{url}

\usepackage{mathtools,calc}

\makeatletter
\newcommand*\concat
  {\mathbin{\mathmakebox[\widthof{${+}\m@th$}]{+\hskip-1emplus1fil+}}}

\makeatother

\title{SC-GlowTTS: an Efficient Zero-Shot Multi-Speaker Text-To-Speech Model}

\name{Edresson Casanova$^1$,  Christopher Shulby$^2$,  Eren Gölge$^3$, Nicolas Michael Müller$^4$,  Frederico Santos de Oliveira$^5$, Arnaldo Candido Junior$^6$, Anderson da Silva Soares$^5$, Sandra Maria Aluisio$^1$,  Moacir Antonelli Ponti$^1$}
\address{\normalsize
    $^1$ Instituto de Ciências Matemáticas e de Computação, University of S\~ao Paulo, S\~ao Carlos/SP, Brazil \\
    $^2$ DefinedCrowd Corp., Seattle, WA, USA \\
    $^3$ Coqui, Berlin, Germany\\
    $^4$ Fraunhofer AISEC, Garching near Munich, Germany \\
    $^5$ Federal University of Goias, Goi\^ania, GO, Brazil \\
    $^6$ Federal University of Technology – Paraná, Medianeira, PR, Brazil
  }
\email{edresson@usp.br}

\begin{document}
\maketitle

\begin{abstract}
In this paper, we propose SC-GlowTTS: an efficient zero-shot multi-speaker text-to-speech model that improves similarity for speakers unseen during training. We propose a speaker-conditional architecture that explores a flow-based decoder that works in a zero-shot scenario. As text encoders, we explore a dilated residual convolutional-based encoder, gated convolutional-based encoder, and transformer-based encoder. Additionally, we have shown that adjusting a GAN-based vocoder for the spectrograms predicted by the TTS model on the training dataset can significantly improve the similarity and speech quality for new speakers. Our model converges using only 11 speakers, reaching state-of-the-art results for similarity with new speakers, as well as high speech quality.
\end{abstract}

\noindent\textbf{Index Terms}: zero-shot multi-speaker TTS, text-to-speech, multi-speaker modeling, zero-shot voice conversion.

\section{Introduction}

Text-to-Speech (TTS) systems have received a lot of attention in recent years due to the great advances by deep learning, which have allowed for the popularization of voice applications such as virtual assistants. Most TTS systems were tailored from a single speaker voice, but there is current interest in synthesizing voices for new speakers, not seen during training, employing only a few seconds of speech samples. This approach is called zero-shot multi-speaker TTS (ZS-TTS) as in \citep{arik2018neural, jia2018transfer, cooper2020zero, choi2020attentron}.

 
ZS-TTS was first proposed~\cite{arik2018neural} by extending the DeepVoice~3~\citep{deepvoice3}. Also, \cite{jia2018transfer} explored Tacotron 2 \citep{tacotron2} using external embeddings extracted from a trained speaker encoder using a generalized end-to-end loss (GE2E)~\citep{ge2e}, resulting in a model that can generate speech, resembling the target speaker. Similarly, \cite{cooper2020zero} explored Tacotron 2 with different speaker embeddings methods. The authors showed that LDE~\citep{cai2018exploring} embeddings improved the similarity and synthesized a more natural speech for novel speakers when compared to X-vector \citep{snyder2018x} embeddings. The authors in \cite{cooper2020zero} also showed that training a gender-dependent model improves the similarity for unseen speakers.


In this context, a major issue is the similarity gap between observed and unobserved speakers during training. In an attempt to reduce this gap, Attentron~\cite{choi2020attentron} proposed a fine-grained encoder with an attention mechanism for extracting detailed styles from various reference samples and a coarse-grained encoder. As a result of using several reference samples instead of one, they achieved a better similarity for unseen speakers.


Despite the recent results, zero-shot multi-speaker TTS remains an open problem in particular concerning the difference in the quality of seen and unseen speakers. Also, current approaches rely heavily on Tacotron 2, while there is potential to improve results with the use of flow-based methods~\citep{kingma2016improved}. In this context, FlowTron~\cite{valle2020flowtron} allowed for the manipulation of multiple aspects of speech, such as pitch, tone, speech rate, cadence, and accent. Also, \cite{kim2020glow} proposed GlowTTS reaching similar quality to Tacotron 2 but with an increase in speed of 15.7 times while permitting speech velocity manipulation.


In this paper, we propose a novel method, Speaker Conditional GlowTTS (SC-GlowTTS), for zero-shot learning of unseen speakers. Our model relies on GlowTTS~\cite{kim2020glow} for the part that converts input characters to spectrograms. SC-GlowTTS uses an external speaker encoder based on Angular Prototypical loss \citep{chung2020in}, to learn speaker embedding vectors, and adapts the HiFi-GAN  \citep{kong2020hifi} vocoder to convert the output spectrograms to the waveform. 
Our contribution is as follows:
\begin{itemize}
    \item A novel zero-shot multi-speaker TTS approach that achieves state-of-the-art results with just 11 speakers in the training set;
    \item An architecture that enables high quality and faster than real-time speech synthesis in the zero-shot multi-speaker TTS setting;
    \item Adjusting a GAN-based vocoder for the spectrograms predicted by the TTS model on the training dataset, in order to significantly improve the similarity and speech quality for new speakers.
\end{itemize}

The audio samples for each of our experiments are available on the demo web-site\footnote{\url{https://edresson.github.io/SC-GlowTTS/}}. In addition, for reproducibility the implementation is available at the Coqui TTS\footnote{\url{https://github.com/coqui-ai/TTS}}, and checkpoints of all experiments are available at the Github repository\footnote{\url{https://github.com/Edresson/SC-GlowTTS}}.



\section{Speaker Conditional GlowTTS Model}\label{sec:TTSModel}



Speaker Conditional Glow-TTS (SC-GlowTTS) builds upon GlowTTS, but includes several novel modifications. In addition to the GlowTTS's transformer-based encoder network, we explore a residual dilated convolutional network \citep{vainer2020speedyspeech} and gated convolutional network \citep{dauphin2017language}; to our knowledge, used for the first time in this context. Our convolutional residual encoder is based on~\citep{vainer2020speedyspeech}, however we used the Mish \citep{misra2019mish} instead of ReLU activation function. On the other hand, our gated convolutional network \citep{dauphin2017language} consists of 9 convolutional blocks and each block includes a dropout layer, a 1D convolution, and a layer normalization \citep{ba2016layer}. We use kernel size 5, dilation rate 1, and 192 channels in all convolutional layers. A flow-based decoder is used with the same architecture and configuration as the GlowTTS model. However, to transform it into a zero-shot TTS model, we include speaker embeddings in the affine coupling layers on all 12 decoder blocks. 
We also used the FastSpeech's duration predictor network \citep{ren2019fastspeech} to predict character durations. To capture different speech characteristics of different speakers, we added speaker embeddings to the input of the duration predictor. Finally, the HiFi-GAN \citep{kong2020hifi} is used as a vocoder.

\begin{figure}[]
  \scriptsize
  \centering
 \includegraphics[width=0.45\textwidth]{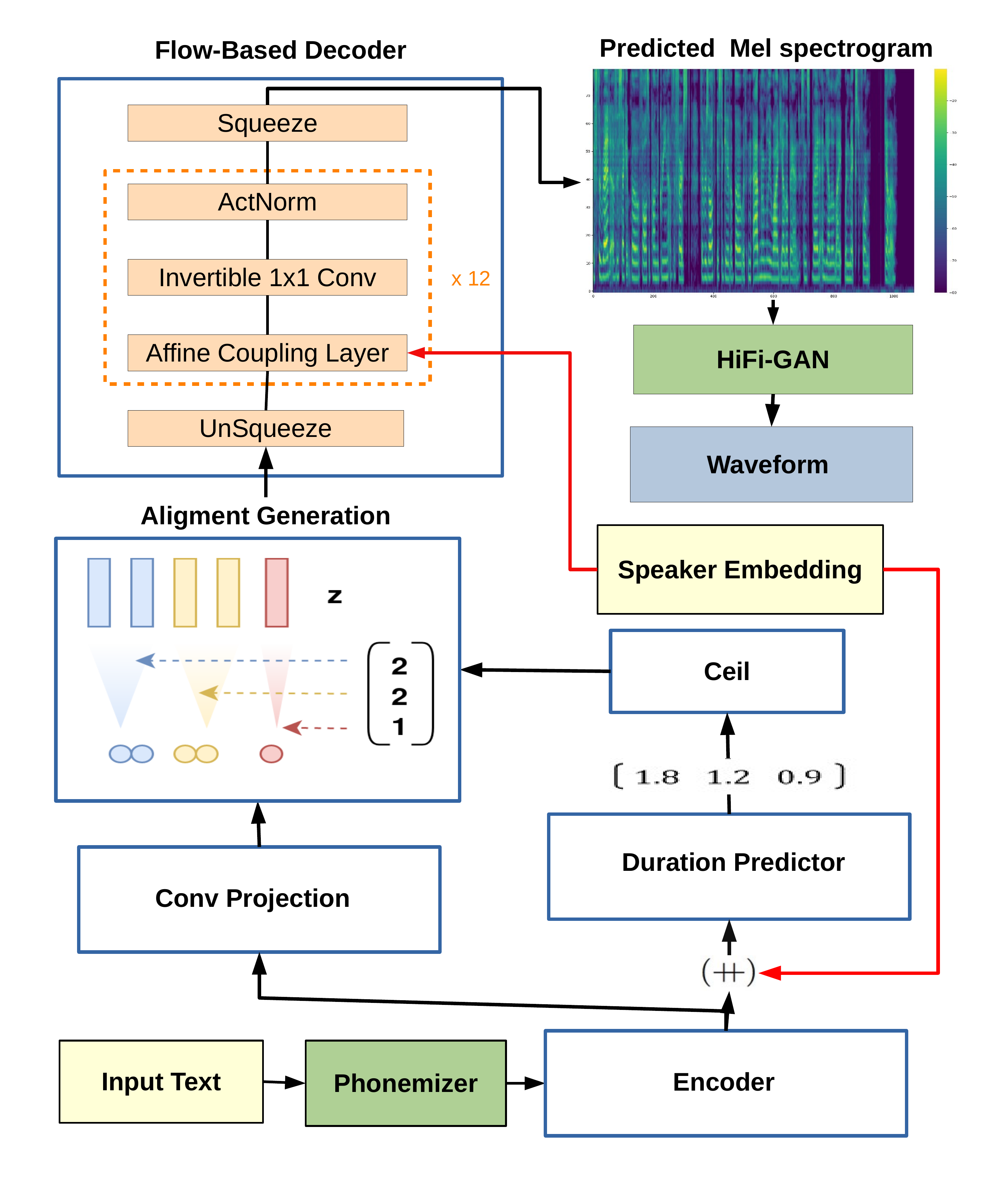}
  \caption{Speaker Conditional GlowTTS General Architecture.}
  \label{fig:generaltop}
\end{figure}

The SC-GlowTTS model, during inference, is illustrated in Figure~\ref{fig:generaltop}, where $(\concat)$ indicates concatenation. During training, the model uses the Monotonic Alignment Search (MAS)~\citep{kim2020glow}, where the decoder's objective is to condition the mel spectrogram and an input speaker embedding in a $P_Z$ prior distribution. The purpose of MAS is to align the $P_Z$ prior distribution with the encoder's output. During inference, MAS is not used, instead, the $P_Z$ prior distribution and alignment are predicted by the text encoder and the duration predictor network. Finally, a latent variable $Z$ is sampled from the prior distribution $P_Z$. Using the inverted decoder and the speaker embeddings, a mel spectrogram is synthesized in parallel, transforming the latent variable $Z$ via the flow-based decoder.

For brevity, we denominate the SC-GlowTTS model with the transformer, residual convolution, and gated convolution based encoders as SC-GlowTTS-Trans, SC-GlowTTS-Res and SC-GlowTTS-Gated model, respectively.

\section{Experiments} \label{sec:method}

\subsection{Speaker Encoder} \label{sec:SE}


Our speaker encoder is a stack of 3 LSTM layers with a linear output layer, similar to \citep{ge2e}. We use 768 LSTM units and 256 units for the linear layer. For training, we used audios sampled at 16 kHz and extracted mel spectograms using a 1024ms window using the Fast Fourier Transform (FFT), with a hop length of 256 and 1024 FFT components, from which we retain only 80 mel coefficients. Optimization was carried out using the Angular Prototypical \citep{chung2020in} loss function different than the original work. The optimizer RAdam \citep{liu2019variance} was used during 320k steps using 64 speakers per batch, with 10 samples of each speaker and a learning rate of $10^{-4}$.

\subsection{Zero-Shot Multi-Speaker Tacotron 2}\label{sec:tacozeroshot}



We compare our approach with Tacotron 2. Following the proposal of  \cite{jia2018transfer}, \cite{paul2020speaker} and \cite{cooper2020zero} we use local sensitive attention \citep{tacotron2}. We concatenate the speaker embeddings to the input of the attention module as in \cite{jia2018transfer,cooper2020zero}, given that the latter showed this was adequate for a gender-independent Tacotron model. To alleviate possible issues in the attention module we use Double Decoder Consistency (DDC) \citep{erogolDDC} with gradual training \citep{erogolgradual} and guided attention \citep{dctts}. In Tacotron, the number of output frames per decoder iteration is called the reduction rate (R) \citep{erogolgradual, tacotron2}. The idea of the DDC is to combine two decoders with different reduction factors. One decoder (coarse) works with a higher R and another decoder (fine) works with a smaller R value. Gradual training simply starts training with a larger R and decreases it during the training. In our experiments, we use $R=7$ for the coarse decoder, while for the fine decoder we used gradual training, starting from $R=7$ and decreasing it as follows: $R=5$ at step 10k; $R=3$ at step 25k; $R=2$ at step 70k.

\subsection{Audio datasets}\label{sec:method:base}
Our speaker encoder was trained with all partitions of the LibriSpeech dataset \citep{librispeech}, the English version of Common Voice \citep{ardila2019common}, the VCTK and VoxCeleb (v1 and v2) datasets \citep{chung2018voxceleb2}, totaling approximately 25k speakers.

Our zero-shot multi-speaker TTS model is trained using VCTK \citep{veaux2016superseded} dataset, an English language dataset containing 44 hours of speech and 109 speakers, sampled at 48KHz. Each speaker pronounces approximately 400 sentences. Pre-processing was carried out in order to remove long periods of silence. We applied voice activity detection (VAD) using Webrtcvad toolkit\footnote{https://github.com/wiseman/py-webrtcvad}. We have divided the VCTK dataset into: train, validation (containing the same speakers as the train set) and test. For the test set, we selected 11 speakers not present in the validation or training set; following the proposal by \cite{jia2018transfer}, we selected 1 representative from each accent totaling 7F/4M (speakers 225, 234, 238, 245, 248, 261, 294, 302, 326, 335 and 347). For the HiFi-GAN \citep{kong2020hifi} vocoder initial training uses \textit{train-clean-100} and \textit{train-clean-360} partitions of the LibriTTS \citep{zen2019libritts} dataset.

\subsection{Experimental setup}\label{sec:method:experiments}

We carried out four training experiments: 

\begin{itemize}
    \item \textbf{Experiment 1:} Tacotron zero-shot model, described in Section \ref{sec:tacozeroshot}, trained for 210k steps.
     \item \textbf{Experiment 2:} SC-GlowTTS-Trans model trained for 150k steps.
    \item \textbf{Experiment 3:}  SC-GlowTTS-Res model trained for 150k steps.
    \item \textbf{Experiment 4:} SC-GlowTTS-Gated model trained for 150k steps.
\end{itemize}

In all experiments, we used RAdam \citep{liu2019variance} with batch size 128, an initial learning rate of $10^{-3}$, and Noam's learning rate schedule~\citep{vaswani2017attention} with 4000 warmup steps. We use the same configuration to extract the mel spectrograms from the speaker encoder, detailed in the Section \ref{sec:SE} but with 22khz sampling rate. We use the VCTK dataset in all our experiments using the training, validation, and test partitions as specified in Section \ref{sec:method:base} and we use the validation set to choose the best checkpoint for each experiment comparing the loss value.


In all experiments, we choose to use phonemes as input instead of text. Specifically, we used the Phonemizer tool\footnote{\url{https://github.com/bootphon/phonemizer/}}, which supports several languages. In addition, we add a blank token between each of the phonemes in the input sentence for the GlowTTS-based models, as suggested by the original work\footnote{\url{https://github.com/jaywalnut310/glow-tts}} \cite{kim2020glow}.


HiFi-GAN v2 model was used as a vocoder, due to its effective speed/quality trade-off. As a starting point, we used the model provided by the authors trained for 500k steps with the LJ Speech \citep{ito2017lj} dataset. We first trained the HiFi-GAN model for 75k steps with the LibriTTS dataset. Afterward, the model is adjusted for other 190k steps using the VCTK dataset, using the training and validation partitions as specified in Section \ref{sec:method:base}.


\cite{kong2020hifi} showed that adjusting the HiFi-GAN model with the spectrogram of the TTS model, improves quality for a single speaker. However, it remains an open question whether it improves; (i) the quality in multi-speaker, (ii) speech similarity for unseen speakers in ZS-TTS settings. To answer this, our TTS models synthesize each of the sentences in the training and validation splits of VCTK dataset. We enabled teacher forcing to keep the alignments between predicted spectrogram frames and the input phonemes. For SC-GlowTTS we use the MAS to align the decoder output with the encoder output. Using these spectrograms extracted from each model, we fine-tuned the checkpoint initially trained with the LibriTTS dataset, for an additional 190k steps, producing the fine-tuned HiFi-GAN (HiFi-GAN-FT).

\section{Results and  Discussion} \label{sec:results}


In this paper, the synthesized speech quality is evaluated using mean opinion score (MOS) study, following \citep{mos}. MOS scores were obtained with rigorous crowdsourcing \citep{dcMOS}. 
For the MOS calculation, 15 professional collaborators per audio were invited from a total of 68 unique contributors (35F/33M). To compare the similarity between the synthesized voice and the original speaker, we calculate the Speaker Encoder Cosine Similarity (SECS). The SECS consists of calculating the cosine similarity between the embeddings of two audios extracted from the speaker encoder. It ranges from -1 to 1, and a larger value indicates a stronger similarity \citep{cooper2020zero}.  Following \cite{choi2020attentron}, we compute SECS using the speaker encoder of the Resemblyzer \citep{ge2e,Jemine2019Master} package; thus, allowing comparison with those studies. 
\textcolor{black}{We also report the MOS similarity (Sim-MOS) following the work of \cite{jia2018transfer} and \citep{choi2020attentron}.}


We also compare the run-time of each model by calculating the Real Time Factor (RTF) on a CPU and GPU. For speed tests we used a machine with an NVIDIA GeForce GTX Titan V GPU, an Intel (R) Xeon (R) CPU E5-2603 v4 @ 1.70GHz processor with 6 CPU cores and 15 Gb of RAM. The training was carried out on an NVIDIA V100 GPU. Also, RTF was calculated considering the full synthesis run, from the input phonemes to the output waveform. We synthesize 15 different sentences as in \cite{peng2020non} 10 times for each of the 11 speakers of the VCTK test set and calculated the average.



As a reference sample for the extraction of speaker embeddings, we use the fifth sentence of the VCTK (i.e, speakerID\_005.txt), since all test speakers uttered it and because it is a long sentence (20 words). In this way, all speakers are presented in the zero-shot multi-speaker TTS model by a reference sample with the same number of words and speech content.

For the calculation of MOS and SECS we randomly drew 55 sentences from the \textit{test-clean} subset of the LibriTTS, considering only sentences with more than 20 words. We randomly select five sentences for each of the 11 test speakers, ensuring that all 55 test sentences are synthesized and that all the test speakers are considered. As ground truth, we select 5 audios randomly for each of the 11 test speakers (55 in total), only audios with more than 20 words are studied.


On the other hand, for the SECS ground truth, we compared the 55 audios chosen at random (explained above) with the reference audios used to synthesize the sentences (fifth sentence of the VCTK dataset for each of the test speakers).

\begin{table*}[]
\centering
\caption{Real Time Factor, MOS and Sim-MOS with 95\% confidence intervals and the SECS for all our experiments.}
\label{tab:rtfsecs}
{\small
\begin{tabular}{l|l|l|l|l|l}
\hline
\textbf{Experiment - Model}               & \textbf{Vocoder}        & \textbf{RTF (CPU - GPU)} &  \textbf{SECS} & \textbf{MOS} & \textbf{Sim-MOS}\\\hline
Ground Truth &-- & -- & 0.9236   & 4.12  $\pm$ 0.06 &  4.127 $\pm$ 0.06 \\\hline
%
\multirow{1}{*}{Attentron ZS \cite{choi2020attentron}} &   WaveRNN         &  --  &     (0.731)   &  (3.86 $\pm$ 0.05)  &  (3.30  $\pm$ 0.06) \\\cline{2-6} \hline
\multirow{2}{*}{1 - Tacotron 2}   
                               & HiFi-GAN         &  0.5782 - 0.2485 &   0.7589       &    3.57 $\pm$ 0.08 &  3.867 $\pm$ 0.08 \\\cline{2-6} 

                               & HiFi-GAN-FT     &   -     &  0.7791      &   3.74   $\pm$ 0.08 &  3.951 $\pm$ 0.07\\\hline
                               
\multirow{2}{*}{2 - SC-GlowTTS-Trans}  
                               & HiFi-GAN         &   0.3612 - 0.1557      & 0.7641           &   3.65 $\pm$ 0.07 & 3.905 $\pm$ 0.07\\\cline{2-6}

                               & HiFi-GAN-FT      &  -       &  \textbf{0.8046}         &   3.78 $\pm$ 0.07 &  \textbf{3.999 $\pm$ 0.07} \\\hline

\multirow{2}{*}{3 - SC-GlowTTS-Res} 
                               & HiFi-GAN         &  0.3597 - 0.1545     & 0.7440         &       3.45 $\pm$ 0.09 & 3.828 $\pm$ 0.08 \\\cline{2-6} 
                               & HiFi-GAN-FT      &  -      & 0.7969   &   3.70   $\pm$  0.07    &   3.916 $\pm$ 0.07 \\\hline 
\multirow{2}{*}{4 - SC-GlowTTS-Gated} 
                               & HiFi-GAN         &  0.3474 - 0.1437            & 0.7432          &       3.55 $\pm$  0.08  & 3.852 $\pm$ 0.08 \\\cline{2-6} 
                               & HiFi-GAN-FT      &  -       & 0.7849     &   \textbf{3.82  $\pm$   0.07} &   3.952 $\pm$ 0.07  \\\hline

\end{tabular}
}
\end{table*}


Table \ref{tab:rtfsecs} shows the RTF in CPU and GPU, MOS with 95\% confidence intervals and SECS for all of our experiments. Speed tests show that the fastest model on both CPU and GPU is SC-GlowTTS-Gated, followed by the SC-GlowTTS-Res model. The SC-GlowTTS-Trans model is the slowest of the SC-GlowTTS family, however, still much faster than Tacotron 2. Despite this, with the integration with the HiFi-GAN vocoder all models are real time in both CPU and GPU.


SECS score of the ground truth reached 0.9222 because it compares the sample used as a reference with other real speech samples of the same speaker. This value is intended to show an upper bound for SECS, i.e., a model that perfectly "copies" the voice of a target speaker.


The best SECS for synthesis with the HiFi-GAN vocoder (without fine-tuning) was obtained by the SC-GlowTTS-Trans model (experiment 2), followed by Tacotron 2 (experiment 1). The SC-GlowTTS-Res model (experiment 3) achieved the third-best SECS being only better than SC-GlowTTS-Gated (experiment 2). Using the HiFi-GAN-FT, the SC-GlowTTS-Trans model also obtained the best SECS, followed by the SC-GlowTTS-Res model. The SC-GlowTTS-Gated model reached the third-best SECS being only better than the Tacotron 2 model. We found that the fine-tuning of the HiFi-GAN vocoder in the spectrograms extracted from the TTS models significantly improves SECS for the new speakers. The SECS increased from 0.7589 to 0.7791, 0.7641 to 0.8046, 0.7440 to 0.7969 and 0.7432 to 0.7849, respectively, for the models Tacotron 2, SC-GlowTTS-Trans, SC-GlowTTS-Res and SC-GlowTTS-Gated. 

\textcolor{black}{For Sim-MOS the results are similar to those of SECS. However, there are some differences which can be explained by the overlapping of the Sim-MOS confidence intervals between the experiments. Improvement with the use of HiFi-GAN-FT can also be seen in all experiments.}


Finally, we compare our results with those presented by the Attentron model. In \cite{choi2020attentron}, the authors reported SECS values, also calculated by the speaker encoder that we use. Although the authors use only 8 speakers (4F/4M) for the test and we use 11 speakers, we believe the comparison is fair and leaves no selection criteria undefined. The Attentron model in zero-shot mode reached a SECS of only 0.731. Such a model uses multiple samples to synthesize speech instead of just one, performing few-shot TTS with 8 reference samples. This approach achieves a SECS of 0.788, slightly lower than our best SECS, 0.8046. Despite the advantage of the few-shot approach, our model still achieves a higher SECS than Attentron. \textcolor{black}{The authors also reported the Sim-MOS reaching 4.83 $\pm$ 0.02 for ground truth speech, zero-shot mode Attentron reaches 3.30 $\pm$ 0.06 and few-shot mode 3.57 $\pm $0.05. Considering these values, our best model was superior at 3,999 $\pm$ 0.07. Furthermore, the results of our model are closer to the ground truth being only 0.128 smaller while in \cite{choi2020attentron} the best experiment's difference is 1.26.
}


For the MOS, ground truth speech reached 4.12. The SC-GlowTTS-Gated model with the HiFi-GAN-FT vocoder was the closest, reaching a MOS of 3.82. Moreover, as in SECS, where the HiFi-GAN-FT vocoder improved speech similarity, the best MOS was achieved using the same vocoder. With the adjustment of the HiFi-GAN vocoder in the spectrograms extracted from the TTS model, the MOS for new speakers increased significantly from 3.57 to 3.74, 3.65 to 3.78, 3.45 to 3.70, 3.55 to 3.82, respectively, for all models Tacotron 2, SC-GlowTTS-Trans, SC-GlowTTS-Res and SC-GlowTTS-Gated. Our MOS values are on par with the other state-of-the-art ZS-TTS models such as \citep{cooper2020zero, choi2020attentron}.

\section{SC-GlowTTS performance with few speakers} 





To emulate a scenario with few speakers, we reflect our test set by selecting a subset of the VCTK dataset training set. This new training set consists of 11 speakers, 7F/4M. We selected 1 representative for each accent, except for the ``New Zealand'' accent that has only one speaker and it is in our test set, so we added an ``American'' speaker instead, the chosen speakers were 229, 249, 293, 313, 301, 374, 304, 316, 251, 297 and 323. \textcolor{black}{From this new training set, we have selected random samples to use as a validation set. As a test set, we use the same one defined in Section \ref{sec:method:base}.}

We use the SC-GlowTTS-Trans model and train it with the LJSpeech~\citep{ito2017lj} dataset for 290k steps. This pre-training in a single-speaker dataset was carried out to prime the encoder of the model in a larger vocabulary. We fine-tuned the SC-GlowTTS-Trans model in the new training set with only 11 speakers for 70k steps and using the validation set, we selected the best checkpoint as step 66k. In addition, using the HiFi-GAN model trained in the LibriTTS dataset for 75k steps, we adjusted for other 95k steps using the same technique. \textcolor{black}{This new experiment resulted in a SECS of 0.7707, a MOS of 3.71 $\pm$ 0.07 and Sim-MOS of 3.93 $\pm$ 0.08. These results are compatible with SECS of 0.7791, MOS 3.74 and Sim-MOS 3.951 $\pm$ 0.07 achieved by Tacotron 2, which used a much larger set of 98 speakers.} Therefore, our SC-GlowTTS-Trans model converges with a 9.8 times smaller dataset, with comparable performance to Tacotron 2. We believe that this is an important step forward especially for ZS-TTS in low-resource languages.

\section{Zero-Shot Voice Conversion}

As in the original GlowTTS \citep{kim2020glow} model, we do not provide any information about the speaker’s identity to the model encoder, so the distribution predicted by the encoder is forced to be independent of the speaker identities. Therefore, like GlowTTS, SC-GlowTTS can convert voices using only the model’s decoder. However, in our work, we condition SC-GlowTTS with external speaker embeddings. It enables our model to resemble the voice for speakers not seen in the training by performing a zero-shot voice conversion. Samples of the zero-shot voice conversion are present on the demo page\footnote{\url{https://edresson.github.io/SC-GlowTTS/}}.

\section{Conclusions and future work} \label{sec:conc}


In this work, we present a novel method, SC-GlowTTS, achieving state-of-the-art ZS-TTS results. We explored three different encoders for the SC-GlowTTS model and showed that a transformer-based encoder gave the best similarity for speakers not seen in the training. Our SC-GlowTTS models are superior to Tacotron 2. Also, when combined with an external speaker encoder, SC-GlowTTS models can perform ZS-TTS with only 11 speakers in the training set. Finally, we found that the adjustment of the HiFi-GAN vocoder in the spectrograms predicted by the TTS model in the training and validation set can significantly improve the similarity and the quality of the synthesized speech (MOS) for speakers not seen in the training. \textcolor{black}{As future work, following the work of \citep{choi2020attentron}, we intend to extend the SC-GlowTTS as a few-shot approach.}
 

\section{Acknowledgements}
This study was financed in part by the Coordena\c{c}\~{a}o de Aperfei\c{c}oamento de Pessoal de N\'{i}vel Superior -- Brasil (CAPES) -- Finance Code 001.  Also, we would like to thank the CyberLabs\footnote{\url{https://cyberlabs.ai/}} for financial support for this paper and DefinedCrowd for making industrial-level MOS testing so easily available. Finally, we would like to thank all contributors to the Coqui TTS repository\footnote{\url{https://github.com/coqui-ai/TTS}}, this work was only possible thanks to the commitment of all.


\bibliographystyle{IEEEtran}
\bibliography{mybib}
\end{document}